# Earlier Web Usage Statistics as Predictors of Later Citation Impact

Tim Brody, Stevan Harnad

**Abstract:** The use of citation counts to assess the impact of research articles is well established. However, the citation impact of an article can only be measured several years after it has been published. As research articles are increasingly accessed through the Web, the number of times an article is downloaded can be instantly recorded and counted. One would expect the number of times an article is read to be related both to the number of times it is cited and to how old the article is. This paper analyses how short-term Web usage impact predicts medium-term citation impact. The physics e-print archive – arXiv.org – is used to test this.

## *Introduction*

Peer-reviewed journal article publication is the primary mode of communication and record of scientific research. Researchers – as authors – write articles that contain experimental results, theories, reviews, and so on. To relate their findings to previous findings, authors *cite* other articles. Authors cite an article if they (a) know of the article, (b) believe it to be relevant to their own article and (c) believe it to be important enough to cite explicitly (i.e., there is both a relevance and an importance judgment inherent in choosing what to cite). It is probably safe to assume that the majority of citations will be positive, but even negative citations (where an author cites an article only to say it is wrong or to disagree with it) will refer to articles the author judges relevant and important enough to warrant rebuttal. Citations can therefore be used as one measure of the importance and influence of articles, as well as by-proxy the importance of the journals they are published in and the authors that wrote them. The total number of times an article is cited is called its *citation impact*.

The time that it takes from the moment an article is accepted for publication (after peer review) -- until it is (1) published, (2) read by other authors, (3) cited by other authors in their own articles, and then (4) those citing articles are themselves peer-reviewed, revised and published -- may range anywhere from 3 months to 1-2 years or even longer (depending on the field, the publication lag, the accessibility of the journal, and the field's turnaround time for reading and citation). In physics, the "cited half-life" of an article (the point at which it has received half of all the citations it will ever receive) is around 5 years (ISI Journal Citation Reports, which shows most physics-based journals having a cited half-life between 3 and 10 years [[7]]). Although articles may continue to be cited for as long as their contents are relevant (in natural science fields this could be forever), citation counts using the ISI Journal Impact Factor ([[6]]) use only 2 years of publication data in a trade-off between (i) an article being recent enough to be useful for assessment and (ii) allowing sufficient time for it to make its impact felt.

Is it possible to identify the importance of an article earlier in the read-cite cycle, at the point when authors are *accessing* the literature? Now that researchers access and read articles through the Web, every download of an article can be logged. The number of downloads of an article is an indicator of its *usage impact*, which can be measured much earlier in the reading-citing cycle.

This paper uses download and citation data from the UK mirror of arXiv.org -- an archive of full-text articles in physics, mathematics, and computer science that have been self-archived by their authors -- to test whether early usage impact can predict later citation impact. For a time-period of two years of cumulative download and citation data the correlation between download and citation counts is found to be .42 (for *High Energy Physics*, N = 14442). When this overall two-year effect is tested at shorter intervals, it turns out that the asymptotic two-year correlation is already reached by 6 months. (As Web log data are available only from 2000 onwards, to allow a two year window of subsequent data only papers deposited between 2000 and 2002 can be included. The correlation r=.4486 is found for papers deposited in 2000 for all subsequent citations and downloads up to October 2004, i.e. 4 years of data for an article from January 2000 to 3 years of data for an article from December 2000.)

The next section describes the arXiv.org e-print archive and the data used from its UK mirror for this study. We describe how the citation data is constructed in Citebase Search, an autonomous citation index similar to CiteSeer. We introduce the *Usage/Citation Impact Correlator*, a tool for measuring the correlation between article download and citation impact. Using the Correlator tool we have found evidence of a correlation between downloads and citations, and in our conclusions suggest that downloads may be used as an early-days predictor of citation impact.

## *arXiv.org*

ArXiv.org [15] is an online database of self-archived [15] research articles covering physics, mathematics, and computer science. Authors deposit their papers as preprints (before peer review) and postprints (after peer review – both referred to here as “e-prints”) in source format (often Latex), which can be converted by the arXiv.org service into postscript and PDF. In addition to depositing the full-text of the article, authors provide metadata. The metadata include the article title, author list, abstract, and optionally a journal reference (where the article is or will be published). Articles are deposited into “sub-arXivs”, subject categories for which users can receive periodical emails listing the latest additions.

The number of new articles deposited in arXiv is growing at a linear rate. Hence, in the context of all the relevant literature (and assuming that the total number of articles written each year is relatively stable), arXiv’s *total annual coverage* – i.e., the proportion of the total annual published literature in physics, mathematics and computer science that is self-archived in arXiv - is increasing linearly. The sub-areas of arXiv are experiencing varying rates of growth. The High Energy Physics (HEP) sub-area is growing least (because most of the literature in this arXiv subject are already being self-archived), whereas Condensed Matter and Astrophysics are still growing considerably (Figure 1).

In addition to the wide coverage of the HEP sub-arXiv, Citebase’s ability to link references in the HEP field is helped by the addition of the journal reference to arXiv’s records by SLAC/SPIRES [15]. SLAC/SPIRES indexes HEP journal articles, and links those published articles to the arXiv e-print. Where an author cites a published article without providing the arXiv identifier, Citebase can use the data provided indirectly by SLAC/SPIRES to link that citation, hence count it in the citation impact.

With 300,000 articles self-archived over 12 years, arXiv is the largest Open Access (i.e., toll-free, full-text, online and web crawler access) *centralised* e-print archive. (There exist bigger archives, such as citeseer, whose contents are computationally harvested from *distributed* sites, rather than being self-archived centrally by their authors, and High-Wire press who provide “free but not open” access). ArXiv is an essential resource for research physicists, receiving about 10,000 downloads per hour on the main mirror site alone (there are a dozen mirror sites). Over the lifetime of arXiv there is evidence that physicist’s citing behaviour has changed, probably as an effect of arXiv’s rapid dissemination model. Figure 3 shows that the latency between an article being deposited and later cited has reduced, from a peak at 12 months (for articles deposited in 1992) to there now being no delay to the peak rate of citations. While the advent and growth of electronic publishing has reduced the time between an author submitting a pre-print and the post-print being published, the evidence from arXiv is that authors are increasingly citing very recent work that is yet to be published in a peer-reviewed journal. This poses an interesting question for what role peer-review – as a gatekeeper to the literature - plays for arXiv.org authors [[11]]. Regardless of the impact on the scientific process, there’s no doubt the rapid dissemination model of arXiv has decreased the read-cite-read cycle.

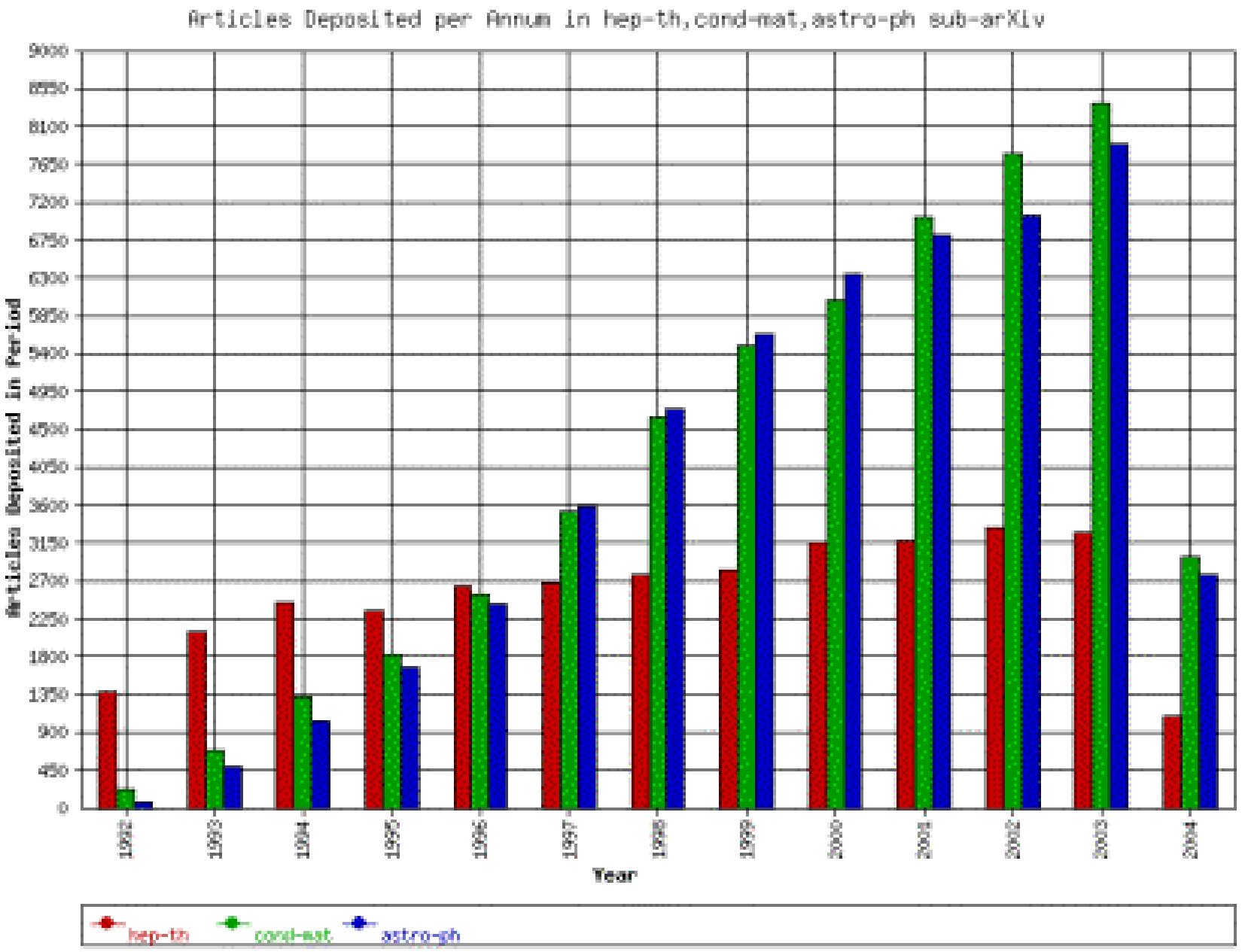

**Figure 1** Deposits in 3 of arXiv.org’s sub-fields. HEP-TH (Theoretical High Energy Physics) has seen little annual growth since the mid-90s. In contrast, COND-MAT (Condensed Matter) and ASTRO-PH (Astrophysics) have continued to see considerable increases in their number of author self-archived additions.

## *Harvesting From arXiv.org*

ArXiv provides access to its metadata records through the Open Archives Initiative Protocol for Metadata Harvesting (OAI-PMH) in Dublin Core format. As the full-text is available without restriction, these are harvested by a Web robot (which knows how to retrieve the source and PDF versions from arXiv's Web interface). Both metadata and full-text are stored in a local cache at Southampton.

Web logs in Apache "combined" format are sent from the UK arXiv mirror (also at Southampton) via email and stored locally. Web logs for the other arXiv mirror sites (including the main site in the US) are currently not made available. The Web logs are filtered to remove common search engine robots, although most crawlers are already blocked by arXiv[1]. Requests for full-texts are then extracted, e.g. URLs that contain "/pdf/" for PDF requests. On any given day only one full-text download of an article from one host is counted (so one user who repeatedly downloads the same article will only be counted once per day). This removes problems with repeated requests for the same article, but results in undercounts when more than one user requests an article from a single host or from behind a network proxy. This study does not count reads from shared printed copies, or reads from electronic copies in different distribution channels such as the publisher.

Each full-text request is translated to an arXiv identifier and stored, along with the date and the domain of the requesting host (e.g. "ac.uk"). This corresponds to some 4.7 million unique requests from the period August 1999 (when the UK arXiv.org mirror was set up) to October 2004. (Because only one mirror's logs are available, this biases the requests towards UK hosts, and possibly towards UK-authored articles; but this cannot be tested or corrected unless the logs are made available from other mirrors.)

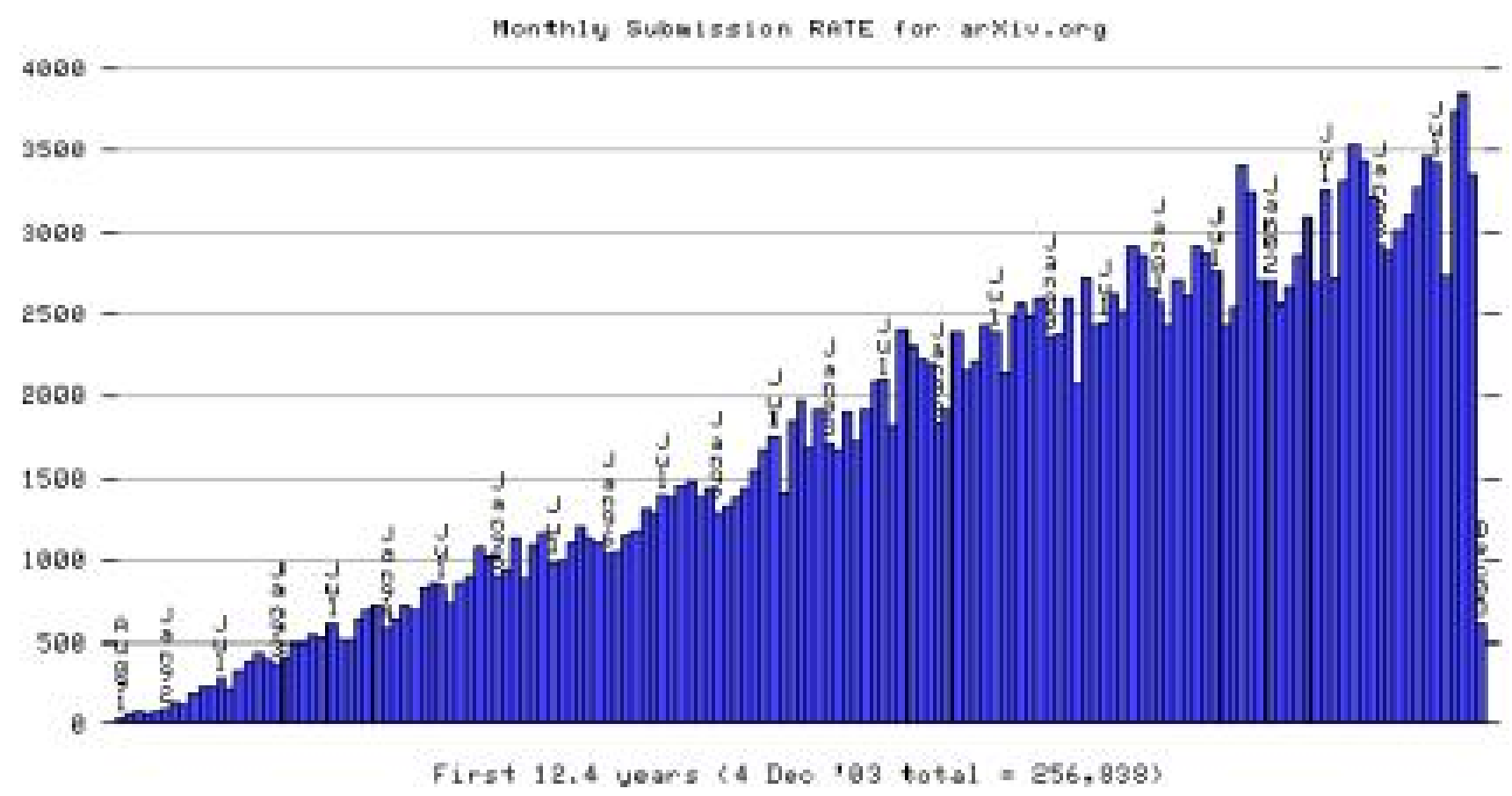

**Figure 2** The monthly number of full-text deposits to arXiv has grown linearly since its creation, to its current level of 4000 deposits per month. (Graph from http://arxiv.org/show_monthly_submissions)

## *Citebase*

Citebase is an autonomous citation index. Metadata records harvested from arXiv.org (and other OAI-PMH archives) are indexed by Citebase. The full-texts from arXiv.org are parsed by Citebase to extract the reference list. These reference lists are parsed, and the cited articles looked up in Citebase. Where the cited article is also deposited in arXiv.org, a citation link is created from the citing article to the cited article. These citation links create a citation database that allows users to follow links to cited articles ("outlinks"), and to see what articles have cited the article they are currently viewing ("inlinks"). Citation links are stored as a list of pairs of citing and cited article identifiers. The total number of citation inlinks to an article provides a *citation impact* score for that article. Within Citebase the citation impact – as well as other performance metrics – can be used to rank articles when performing a search.

The citation impact score found by Citebase is therefore dependent upon several systematic factors: whether the cited article has been self-archived, the quality of the bibliographic information for the cited article (e.g. the presence of a journal reference), the extent to which Citebase could parse the references from citing articles, and how well the bibliographic data parsed from a reference matches the bibliographic data of the cited article. Citebase's citation linking is based either upon an arXiv.org identifier (if provided by the citing author), or by bibliographic data. Linking by identifier can lead to false positives, where an author has something in their reference that looks like an identifier but isn't actually, or where an author has made a mistake (in which case the link goes to the wrong paper). Linking by bibliographic data is more robust, as it requires four distinct bibliographic components to match (author or journal title, volume, page and year), but will obviously be subject to some false positives (e.g. where two references are erroneously counted as one) and uncounted citations. No statistical analysis has been performed on Citebase's linking ability, mainly because such a study could only be performed through exhaustive human-verification of reference links (even then, such a study would be subject to human error!).

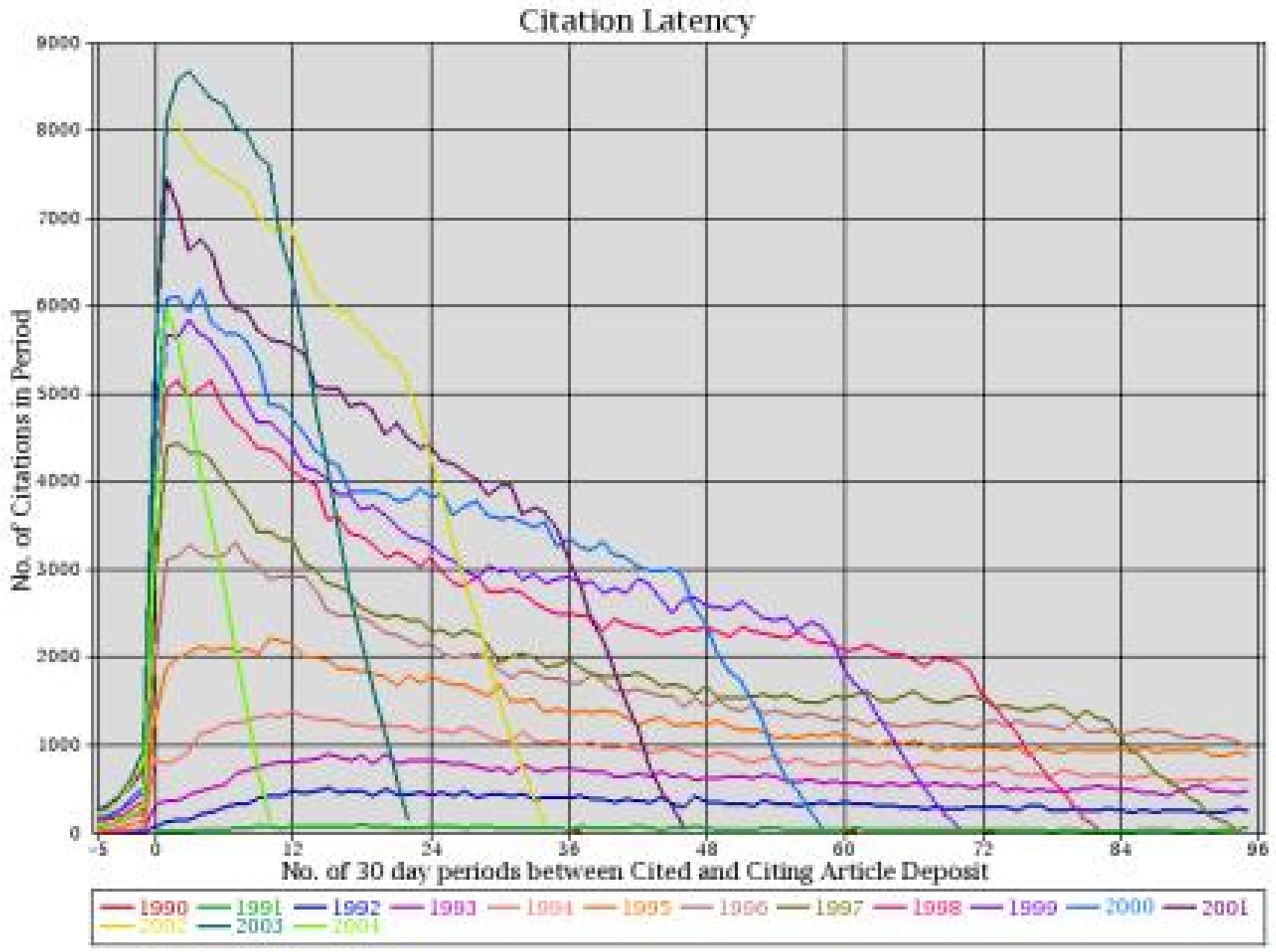

**Figure 3** Citation latency is the number of days between a citing and cited article being deposited (pair-values). This graph plots the frequency of citation latencies by the year the **cited** article was deposited. Each line represents a different sample year, with newer sample years containing more articles hence a higher line on the graph. The significance of this graph lies in the changing distribution of latencies, as e.g. for articles deposited in 1992 the highest rate of citations occurred in the following year (+12 months). The delay before the highest rate of citations has since decreased to seemingly nothing (see also Figure 4). Negative latencies occur when the citing article has an accession date after the cited article. Negative latencies are due to three possible situations; a) an article is updated to include references to new articles (a facility supported by arXiv.org), b) an article cites a published version for which the e-print was later deposited, or c) an author has cited an article they know will exist but hasn't yet been published (e.g. they have read a draft).

## *Correlation between Citations and Downloads*

Correlation is a statistical measure of how two variables covary. Two positively correlated variables x and y will tend to have high values of x paired with high values of y, and low values of x with low values of y. A negative correlation is where high values of x are paired with low values of y. Correlation is a normalised, scale-independent measure based on standard deviations above and below each variable's mean - the raw values of x and y can be in any number range. A correlation between x and y may occur because x influences y, y influences x, the influence is in both directions, or a third variable influences both x and y. Intuitively, one would expect citations and downloads to exert a bi-directional influence, cyclical in time: An author reads a paper A, finds it useful and cites it in a new paper B (download causes citation). Another author reads B, follows the citation, reads A (citation causes download) and then perhaps goes on to cite it in another paper, C (download causes citation), etc. The correlation will be less than 1.0, not only because we don't cite everything we read, nor read everything that a paper we read cites, but because both downloads and citations are subject to other influences outside this read-cite-read cycle (e.g., from alternative discovery tools, or when authors copy citations from papers they read without actually reading the cited works). We would expect reader-only users to reinforce the cite-read influence (as users follow citations), but not to increase the cite-read influence as they do not contribute citations to the system.
Monitoring the correlation between citations and downloads is also informative because although papers can be downloaded and cited for as long as they are available, the peak rate of downloads and citations tend to occur at different time periods. Articles in arXiv.org that are over a year old show an almost flat rate of downloads whereas their citation rate shows a more linear rate of decay over the period of available data [Figure 4]. If there is a correlation between citations and downloads, a higher rate of downloads in the first year of an article could predict a higher number of eventual citations later.

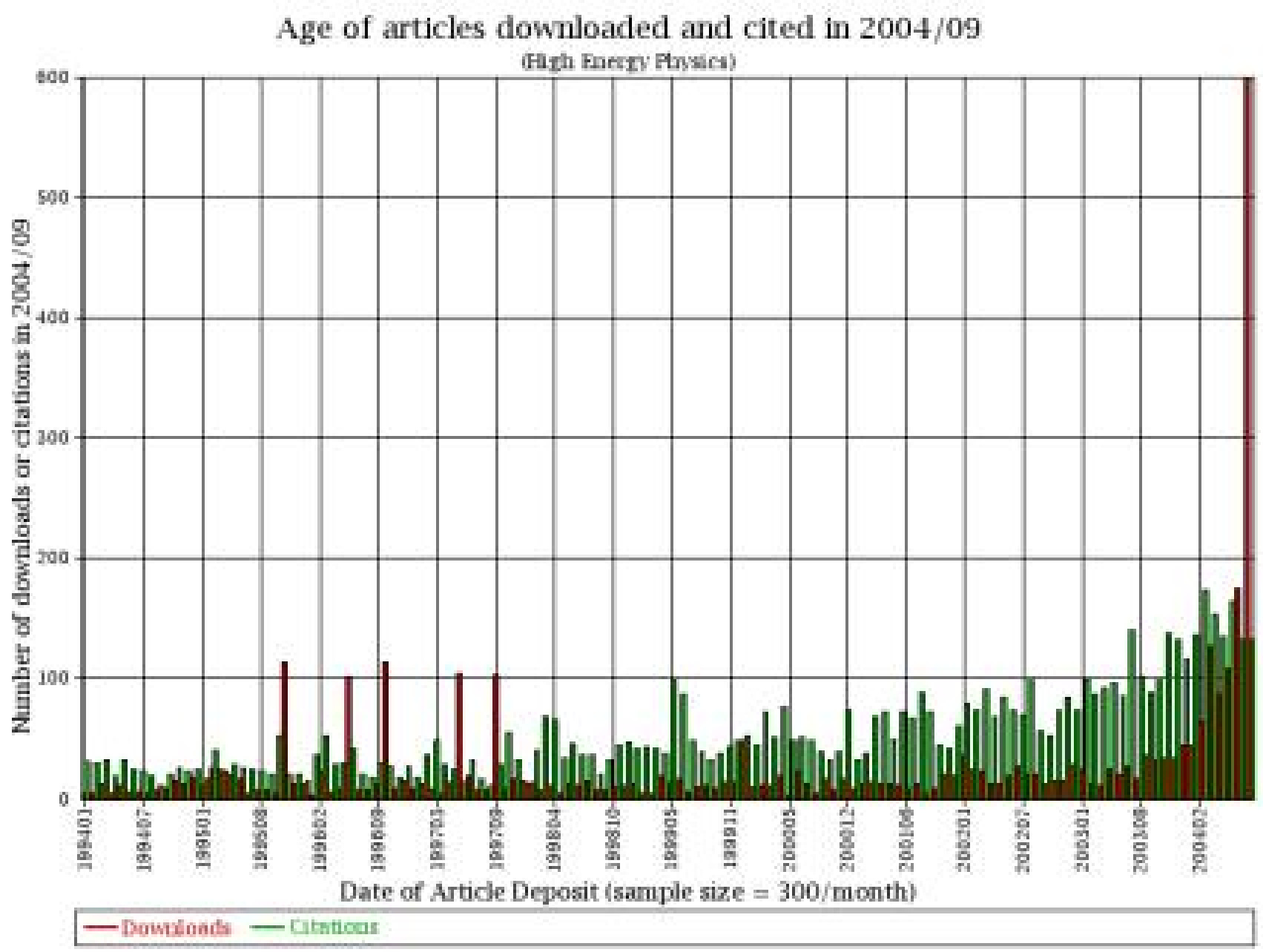

**Figure 4** Temporal properties of downloads and citations. The most frequently downloaded articles are those deposited in the previous year, with the likelihood of articles older than a year downloaded being fairly even. The distribution of the age of cited articles is noticeably different, with a slight decreasing curve of articles up to 6 years old and then - similarly to downloads - an approximately flat distribution over age. The data are downloads in 2004-09 and citations from articles deposited in 2004-09. A random sample of 300 articles were chosen from each preceding month until 1994-01, with the number of downloads and citations to each month's sample plotted as a bar (downloads as red, citations as green).

We built a "Correlation Generator" to analyse the relationship between the citation and download counts for research articles in arXiv.org and to test whether a higher rate of downloads leads to a higher rate of citations.

### *Correlation Generator*

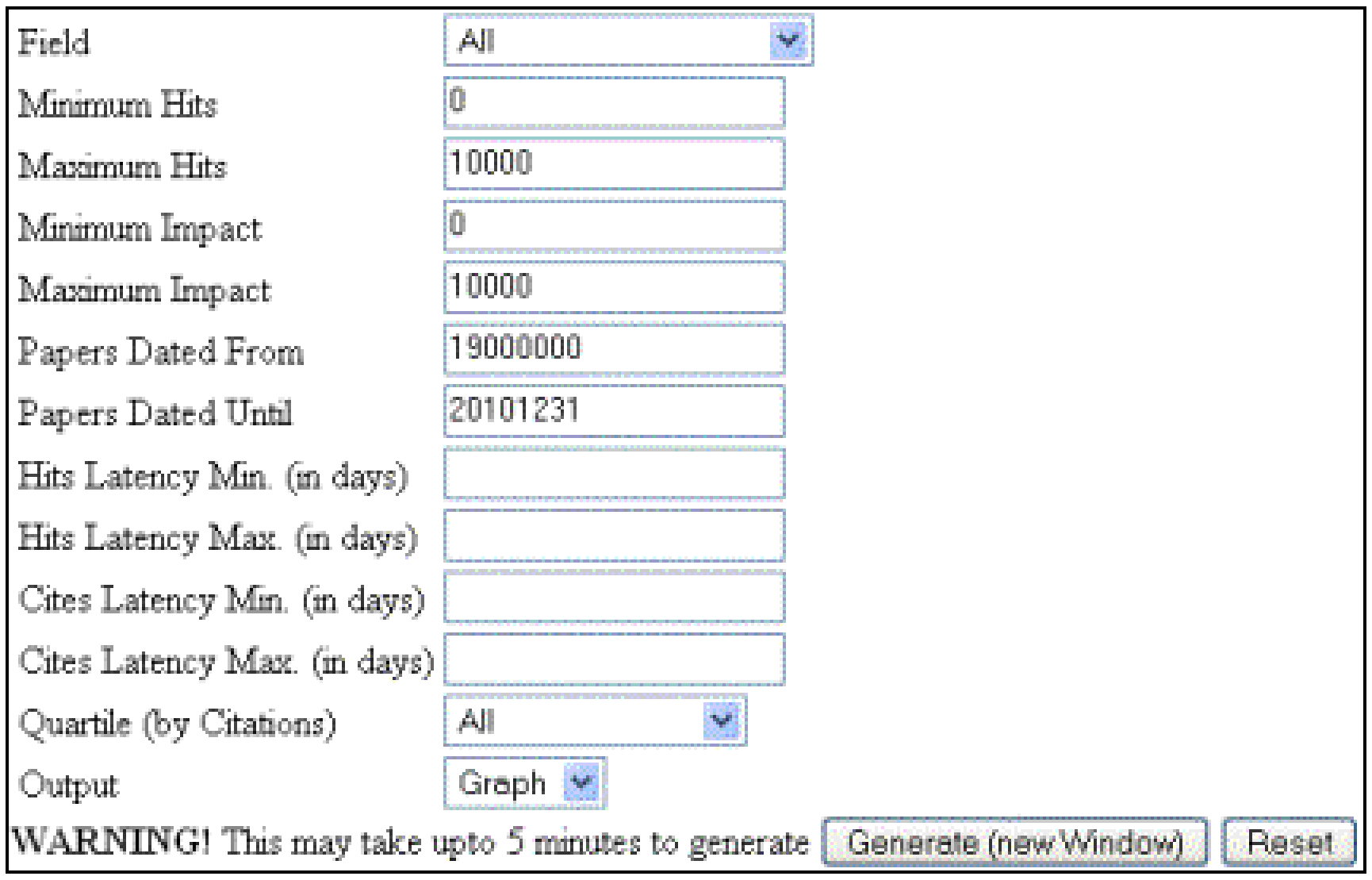

**Figure 5** Form used to generate correlations. This allows the user to choose what data and data-ranges are used to generate the correlation between downloads ("downloads") and citations. This provides filters to delimit which papers (Field, Min/Max Downloads, Min/Max Impact, Date), which downloads and citations (Min/Max downloads latency, Min/Max citation latency), and which citations quartile to include. The citation quartile includes in the result only the bottom, lower, upper, or top 25% of papers after rank-ordering by citation impact. Clicking 'Generate' calls a Web script that extracts the data sets from Citebase, generates the correlation, and displays the result as a graph.

The correlation generator provides a number of filters that can be used to restrict the data going into the correlation. As the correlation is calculated from pairs of citation and download counts corresponding to one article, the filters determine which articles to include and which downloads and citations to count. The values used can be specified symbolically, by entering values into the form, or iconically, by clicking the upper and lower limits on the mini-graphs relevant to the filter (Figure 6).

Articles can be filtered in terms of their arXiv sub-area (e.g. High Energy Physics), the date the article was deposited, and the total number of citations/downloads to each article. This is particularly useful for restricting the analysis to articles for which sufficient data ate available: For example, whereas there exist articles that have been deposited since 1991, the download data are only available from 1999. Hence although the download data cover all of the articles deposited up to that date, the predictive power of downloads can only be tested for articles deposited since 1999.
Each citation and download has a latency value - the time between the cited and citing article being deposited, and the time between the article being deposited and later downloaded. The user might chose to include only downloads that occurred up to a week ("7 days") after an article was deposited.
(don't think I need to repeat this)
Once all the filtered pairs have been found, the natural logarithm is calculated for both values, to allow a generalised linear correlation to be performed.

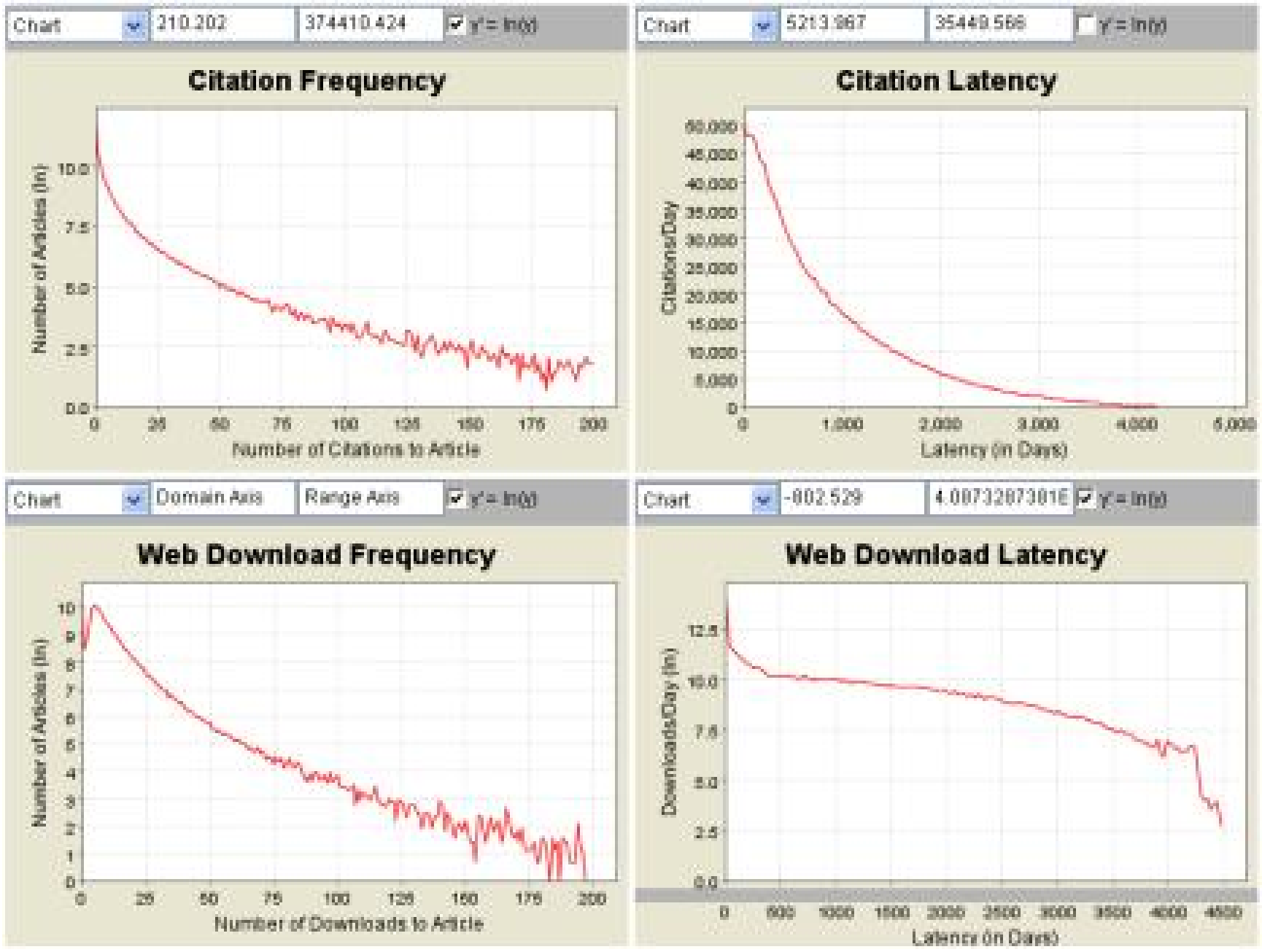

[Figure ](#)**6** Auxiliary graphs. These graphs show the distribution of the variables that go into the correlation (note that most use a **logarithmic scale**). Citation frequency shows the distribution of articles in terms of the number of times they were cited. Citation latency is the time between an article being deposited and later cited (total citations per day latency). Web download frequency is the distribution of articles in terms of the number of times they were downloaded. Web download latency is the time between an article being deposited and later downloaded. The user can click on the graphs to set minimum and maximum values, which are filled into the query form (Figure 5).

## *Sample Correlations*

The correlation generator builds a scatter graph, as well as calculating the basic distribution of the citation and download counts, and the correlation between the two. The scatter graph is made of density dots – the darker the colour the more pairs exist at the same point. This helps to emphasize where the bulk of the pairs lie.
The basic statistical information on the number of pairs used (n), and the distributions of the two variables (sum, mean, and standard deviation) is shown. Both citations and downloads have some large deviations from the mean; this reflects the fact that a small number of very-high-impact papers account capture most of the downloads and citations, while the majority of papers receive few or no downloads or citations. (Figure 6 illustrates that the distributions for downloads and citations show exponential decay. ) The correlation is based on the natural logarithm of the two input variables; hence it is displayed on the axis using the ln() function (Figure 7).
Citation and download frequencies are discrete (i.e. an article can't have 4.5 citations), hence on a logarithmic scale obvious lines of points occur at 0, 1, 2, 3, etc. values. The overall correlation of ~0.4 is reflected in the distribution of pairs away from the line of best fit. However, it can be seen that the overall distribution of pairs becomes more highly correlated towards higher values of citations with higher values of downloads. In particular, there are few papers that have high numbers of citations, but very few downloads. That there are papers with high numbers of downloads associated with low numbers of citations might just reflect a bias against articles for which it has proved technically difficult to find the citation links (commonly where the author hasn't supplied the journal reference, hence any citations to those articles using only a journal reference cannot be linked and counted).

| Un-normalised | Normalised (ln()), approx. |
|---|---|
| 1 | 0 |
| 10 | 2.3 |
| 100 | 4.6 |
| 1000 | 6.9 |
| 10000 | 9.2 |

**Table 1** Table of powers of 10 to their logarithmic values (provided for reference). Values on scatter graph axis are logarithmic.

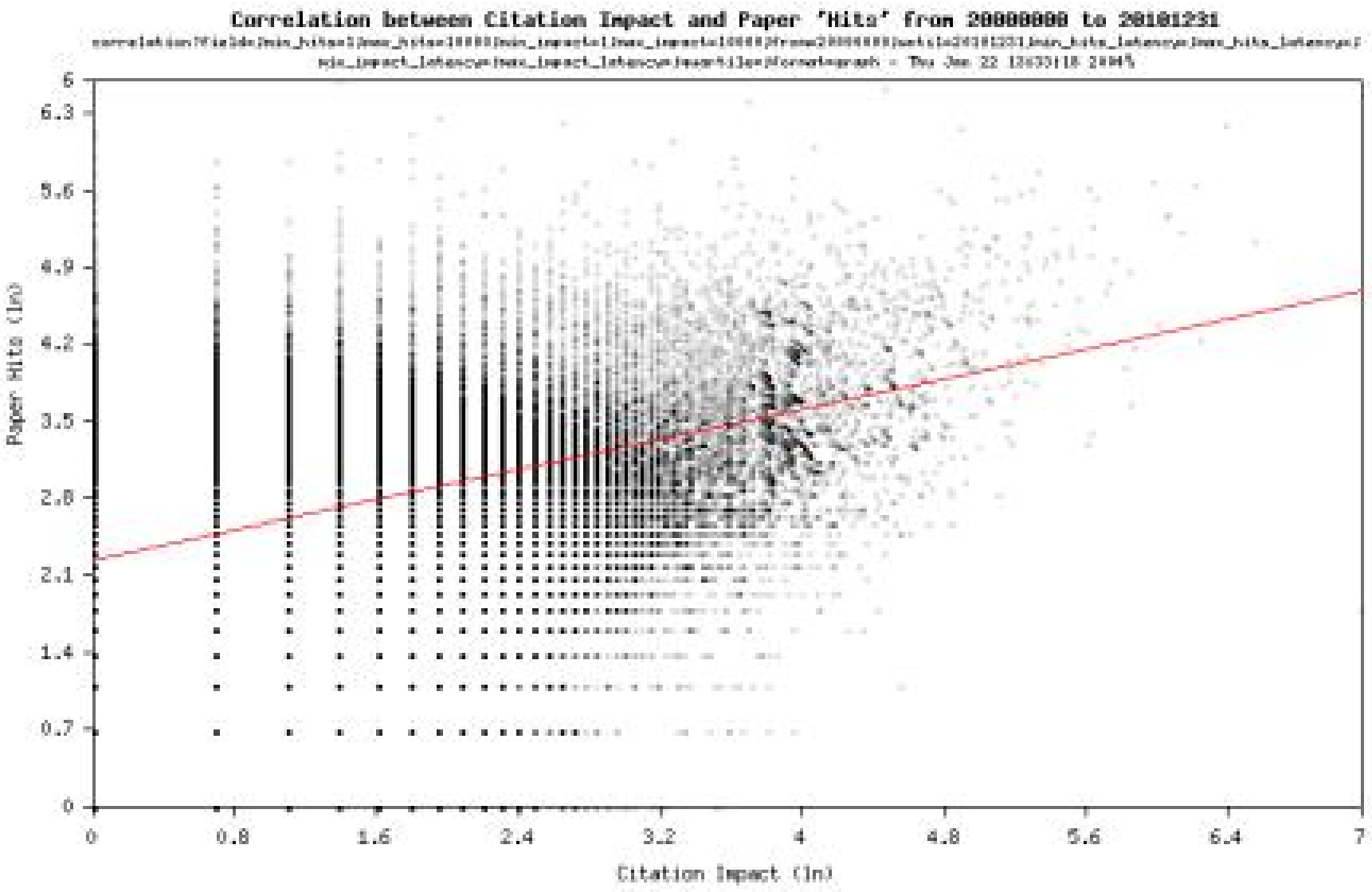

| | Min Filter | Max Filter | Sum (0.43) | Mean | Standard Dev. |
|---|---|---|---|---|---|
| Impact | 1 | 10000 | 83400.23 | 1.15 | 1.07 |
| Hits | 1 | 10000 | 190160.63 | 2.63 | 0.89 |

| | |
|---|---|
| r (squared) | 0.3432 (0.1177) |
| n (t) | 72279 (98.2384) |
| Non-Dir. Sig. | <0.0001 |

**Figure 7** Correlation scatter-graph generated for all papers deposited between 2000-2004[2]. Each dot corresponds to an article. The number of articles with the same values is indicated by shades of grey (black being the highest, with 4 or more articles having the same number of downloads and citations). The download and citation counts are the cumulative amounts up to January 2004. The correlation for these 72,279 papers is r=0.3432. From the distribution in the scatter graph it can be seen that the distribution is very noisy, but that few articles with high citation impact receive low download impact. The ratio of downloads to citations is 2.28:1 (only download statistics for the UK mirror are available), which corresponds to a mean of 1.15 citations for each article, and 2.63 downloads.

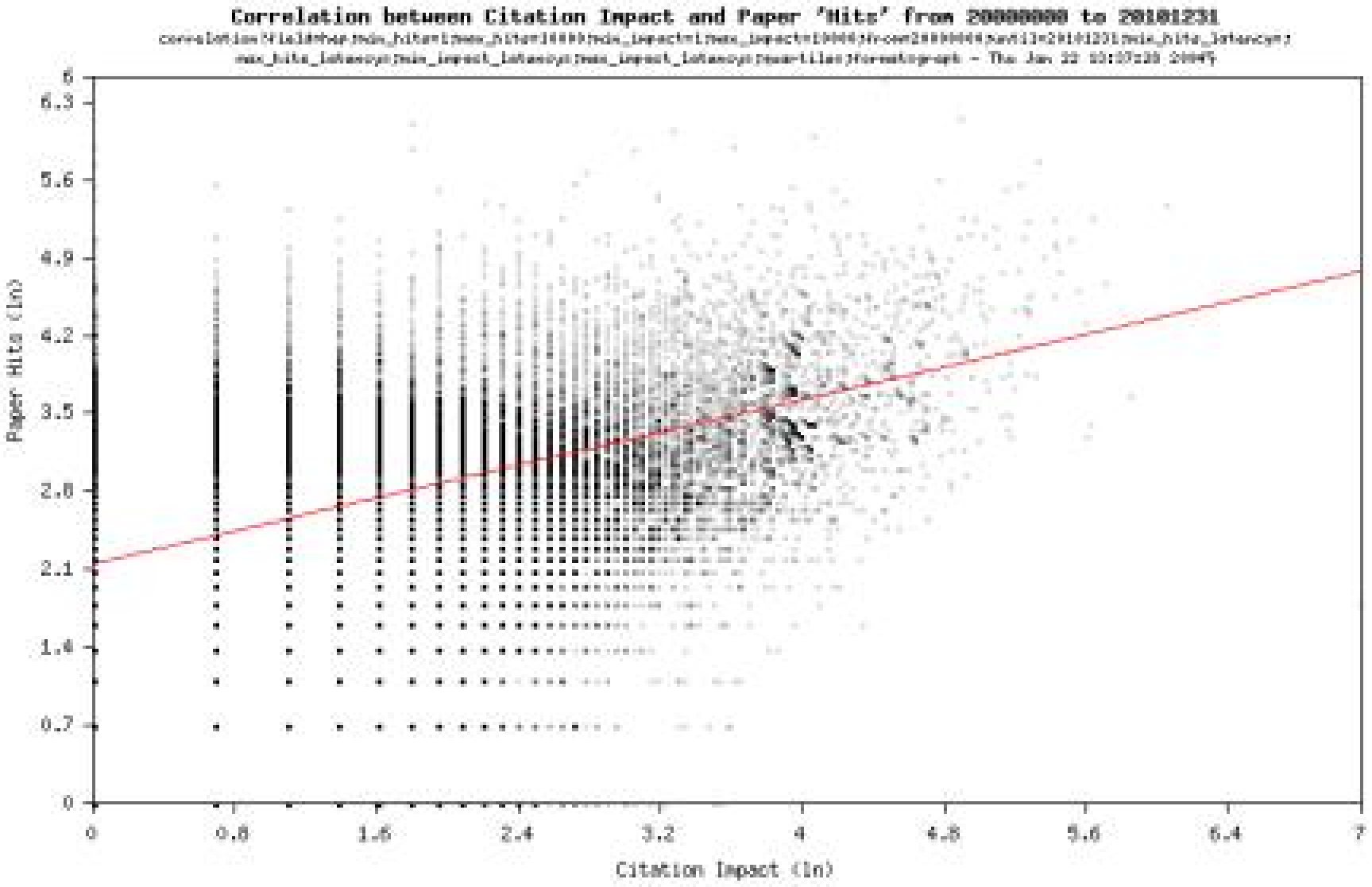

| | Min Filter | Max Filter | Sum (0.56) | Mean | Standard Dev. |
|---|---|---|---|---|---|
| Impact | 1 | 10000 | 39318.56 | 1.54 | 1.15 |
| Hits | 1 | 10000 | 69358.49 | 2.72 | 0.88 |

| | |
|---|---|
| r (squared) | 0.3675 (0.135) |
| n (t) | 25467 (63.0688) |
| Non-Dir. Sig. | <0.0001 |

**Figure 8** Correlation of papers deposited in 2000-current, from the High Energy Physics sub-arXivs[3]. Compared to the overall arXiv, HEP papers have a mean citation impact of 1.54 (1.15), and download impact of 2.72 (2.63). The correlation is higher at r=0.3675.

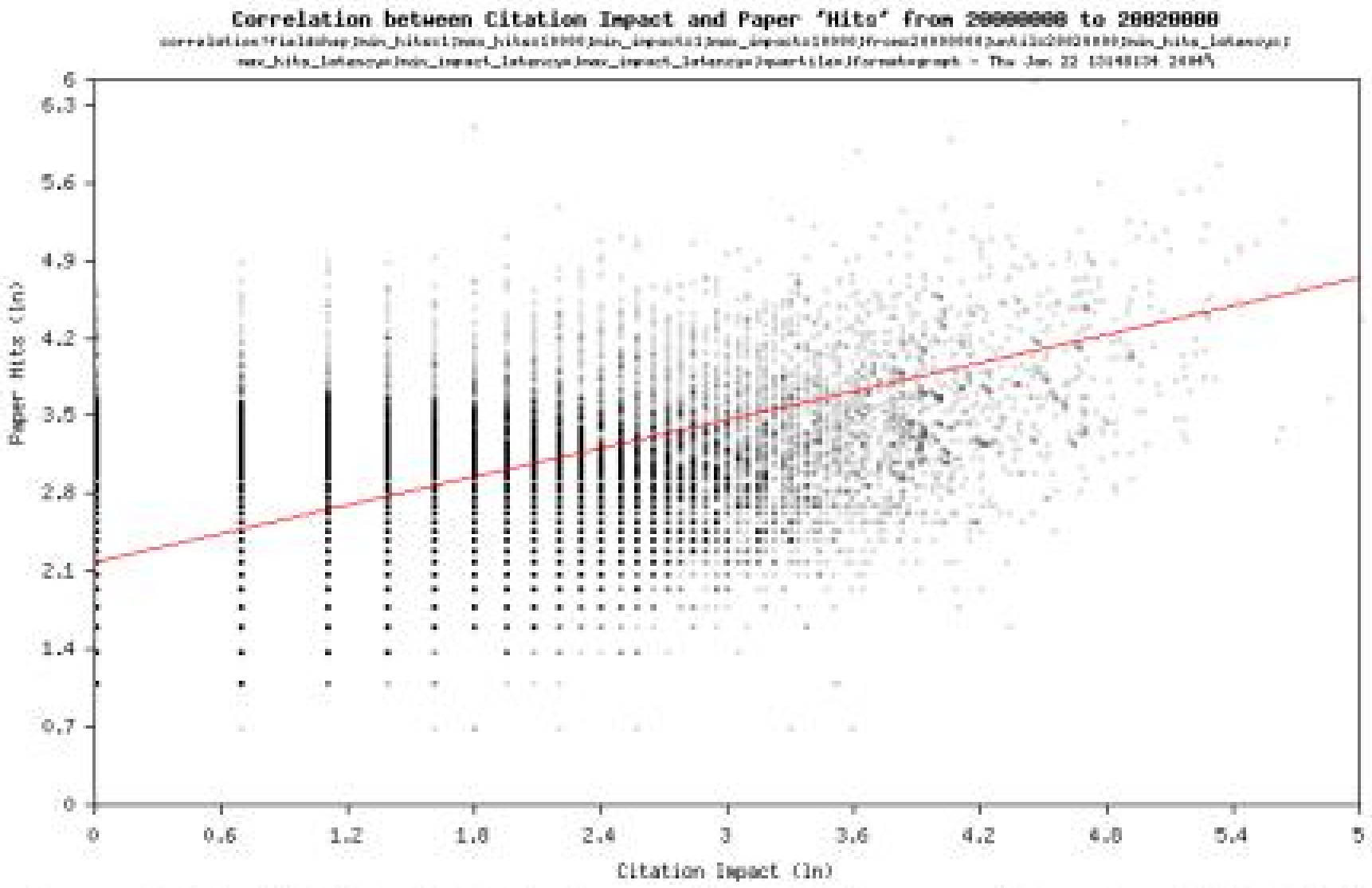

| | Min Filter | Max Filter | Sum (0.6) | Mean | Standard Dev. |
|---|---|---|---|---|---|
| Impact | 1 | 10000 | 25561.31 | 1.77 | 1.18 |
| Hits | 1 | 10000 | 42347.11 | 2.93 | 0.67 |

| | |
|---|---|
| r (squared) | 0.4281 (0.1832) |
| n (t) | 14442 (56.9264) |
| Non-Dir. Sig. | <0.0001 |

**Figure 9** Correlation of papers deposited 2000-2002 (a 2 year period), in the HEP fields3. Given that there is a delay in citation impact differentiating papers, restricting the sample set to papers deposited more than 2 years ago provides a better correlation (0.4281).

## *Predicting from Correlations*

The papers used for testing how well downloads can be used to predict citation impact are from the High Energy Physics sub-arXivs3 and deposited between 2000 and 2002 (a 2 year period). In order to normalise the data, papers with no downloads or no citations are excluded, as the natural logarithm of 0 is undefined. The first 7 days of downloads are excluded, to minimise the effect of the rush of downloads where all paper-titles are read equally after appearing in alert lists.

To test how well downloads predict citation impact, different latency filters are used for the download impact e.g. “How well does the download impact after 2 months predict the citation latency after 2 years?” The correlation generator allows the user to ask this question by specifying the maximum number of days for which to include downloads and citations after the paper is deposited. We will treat a 30 day period as a ‘month’, and 730 days as 2 ‘years’.

Given that citation and download impact have an overall correlation of at least 0.4, how soon can download impact be used to predict citation impact? To test this, queries were made to the correlation generator using 9 different time periods for download data: 1, 2, 3, 4, 5, 6, 7, 12 and 24 months following the deposit of a paper (Table 2). The correlation increases from 0.2712 at 1 month of download data to 0.4321 at 24 months. Figure 11 reveals and important finding: this increase is not linear, and it approximates the final correlation after 6-7 months. This suggests that if the baseline correlation for a field is significant and sufficiently large, the download data could be used after 6 months as a good predictor of citation impact after 2 years.

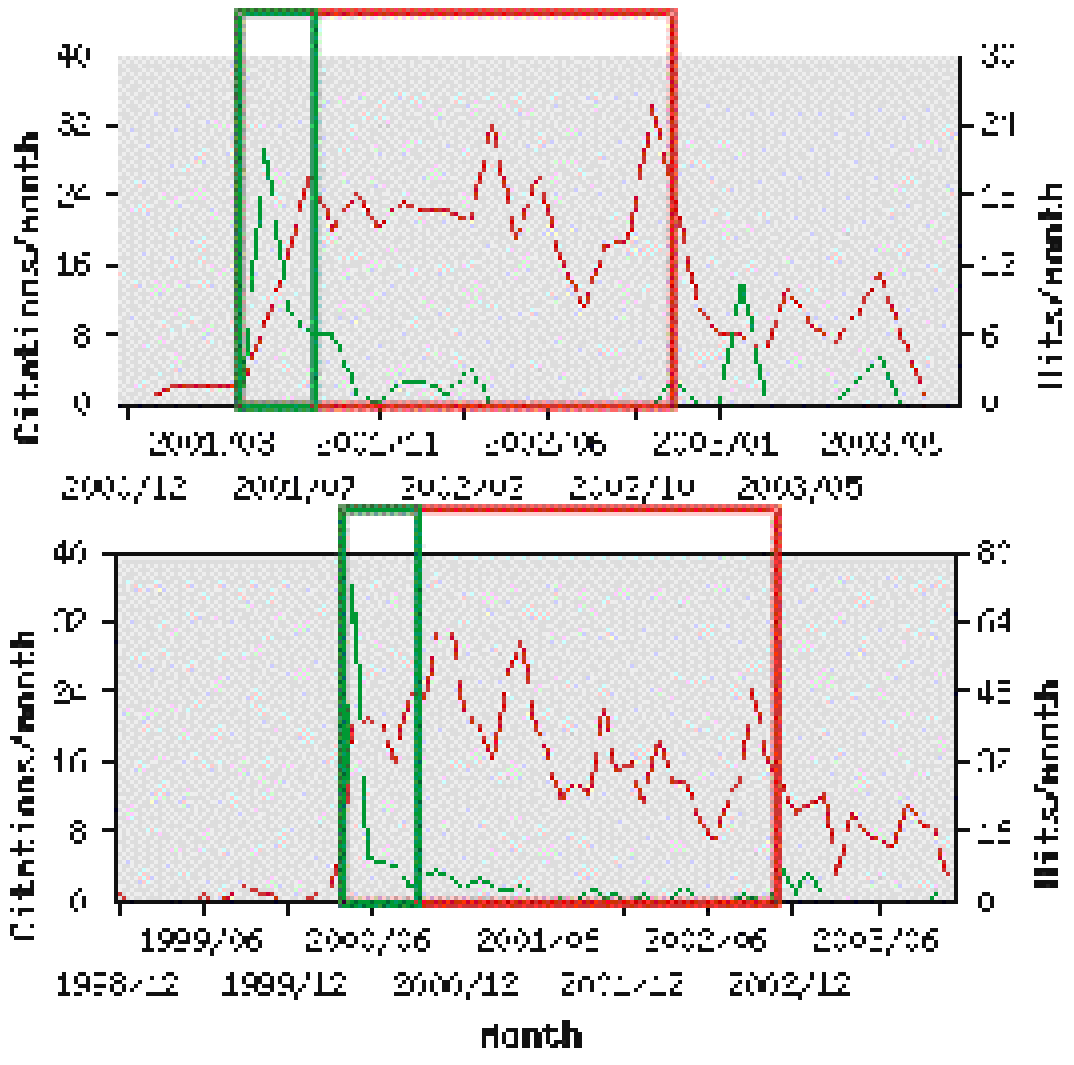

**Figure 10** These graphs show downloads (green) and citations (red) that would be included in calculating the correlation for two papers, where downloads were included up to 4 months after deposit and citations up to 2 years. Note: while these papers were deposited in 2001/04 and 1999/10, respectively, some articles that cite them were deposited earlier and may have been citation-linked using a journal citation or may have been updated to include new citations. The citation latency is taken as the time between the **first** deposit in arXiv.org of both papers (i.e. updates have no bearing on the citation latency, even though author-updates are often to add new citations as indicated by arXiv.org's "comment" field).

| Max. Download Latency (days) | Mean Downloads/Paper (excl. first 7 days) | Correlation (r) |
|---|---|---|
| [30]() | 0.85 | 0.2712 |
| 60 | 1.10 | 0.3239 |
| 90 | 1.26 | 0.3533 |
| 120 | 1.39 | 0.3699 |
| 150 | 1.49 | 0.3829 |
| 180 | 1.57 | 0.3930 |
| 210 | 1.64 | 0.3982 |
| *365* | *1.86* | *0.4177* |
| *730* | *2.20* | *0.4321* |

[Table ]() **2** Correlation between citation and download impact at different maximum download latency periods. The longer the period for which downloads are counted, the higher the correlation between citation and download impact. After 6 months the correlation only increases by a small amount, suggesting that counting the download to an article at 6 months will provide as good an indicator of the citation impact of a paper after 2 years as counting the downloads after 2 years.

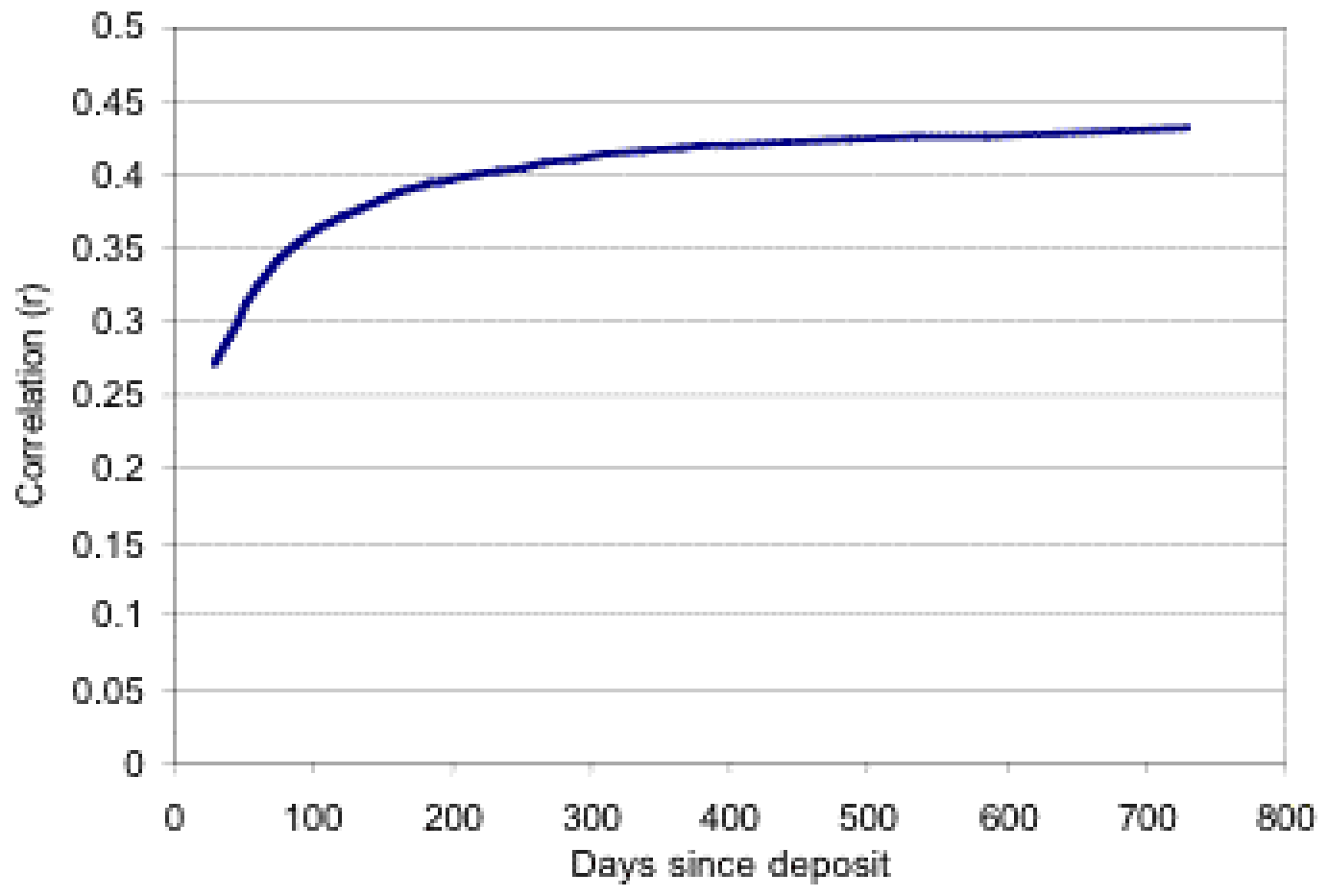

[Figure ]() **11** How correlation varies with how long downloads are counted for.

## *Conclusion*

[tC1]Whereas the significance of citation impact is well established, access of research literature via the Web provides a

new metric for measuring the impact of articles – Web download impact. Download impact is useful for at least two reasons: (1) The portion of download variance that is correlated with citation counts provides an early-days estimate of probable citation impact that can begin to be tracked from the instant an article is made Open Access and that already attains its maximum predictive power after 6 months. (2) The portion of download variance that is uncorrelated with citation counts provides a second, partly independent estimate of the *impact* of an article, sensitive to another form of research usage that is not reflected in citations (Kurtz 2004).
This study found a significant and sizeable correlation of approximately 0.4 between the citation and download impact of articles in physics and mathematics. This was based on Web downloads from the UK arXiv.org mirror only, and on those citations that could be automatically found and linked by Citebase. The true correlation may in fact prove somewhat higher once more download sites are monitored and automatic linking becomes more accurate. It will no doubt vary from field to field, and may also change as the proportion of Open Access content (now 10-20%) approaches 100%.

---

[1] In addition to having a "robots.txt" (that declares full-text downloads off-limits to all sites, with the exception of Google) arXiv watches for rapid downloads from a single site to restricted areas of the Web site, and blocks those sites if they continue after being given a warning.
[2] Web log data is only available since 1999-08.
[3] The hep-th, hep-ph, hep-lat, and hep-ex sub-arXivs are the longest established, hence most comprehensive and well citation-linked parts of arXiv.org.

---

[tC1][A correlation of 0.4 can predict about 16% of the variance. That is good. If it was near 100%, it would not have any chance of having some independent significance of its own. As it is, it can predict 16% of eventual citation variance. Maybe enhanced with some hubs/authorities predictors on citations (**who** cites the paper?), and maybe some co-citation analysis and eventually perhaps even some co-download analysis and hubs/authorities-like weights on

downloads, these can all form a big regression equation predicting other aspects of impact (eventual prizes, funding, research direction, influence, originality, etc.) The results are excellent, and a correlation of .4 is just fine. Try to get an equivalent of Kurtz's 17/1 and 12/1 "reads/cites" ratio (he reports astro) and add references and conclusions and submit for publication! – S.H.]